\documentclass[a4paper]{jpconf}

\usepackage{graphicx,color}
\usepackage{amssymb}
\usepackage{amsmath}

\begin{document}
\title{DMRG analysis of the SDW-CDW crossover region in the  
1D half-filled Hubbard-Holstein model}
\author{S.\ Ejima and H.\ Fehske}
\address{Institut f{\"ur} Physik,
          Ernst-Moritz-Arndt-Universit{\"a}t Greifswald,
          17489 Greifswald,
          Germany}

\date{\today}

\begin{abstract}
In order to clarify the physics of the crossover
from a spin-density-wave (SDW) Mott insulator to a
charge-density-wave (CDW) Peierls insulator in one-dimensional (1D)
systems, we investigate the Hubbard-Holstein Hamiltonian at half 
filling within a density matrix renormalisation group (DMRG) approach. 
Determining the spin and charge correlation exponents, 
the momentum distribution function, and various excitation gaps, 
we confirm that an intervening metallic phase expands the 
SDW-CDW transition in the weak-coupling regime.  
\end{abstract}

The Hubbard-Holstein model (HHM)~\cite{HHM83} is archetypal for exploring 
the complex interplay of electron-electron and electron-phonon 
interactions especially in quasi-1D materials, such as 
halogen-bridged transition metal complexes, charge transfer salts, 
or organic superconductors~\cite{TNYS90}.
It accounts for a tight-binding electron band $(\propto 2t$), 
an intra-site Coulomb repulsion between electrons of opposite spin
$(\propto u=U/4t)$, a local coupling of the charge carriers to optical 
phonons $(\propto \lambda=g^2\omega_0/2t)$, and the energy of the phonon 
subsystem in harmonic approximation $(\propto \omega_0/t$):  
\begin{eqnarray}
 {\cal H}=-t\sum_{j \sigma} 
                (c_{j\sigma}^{\dagger}
                 c_{j+1\sigma}^{\phantom{\dagger}}
                 +{\rm h.c.}
                )
	      +U\sum_j n_{j\uparrow}n_{j\downarrow}
	      -g\omega_0\sum_{j\sigma}
                 (b_j^{\dagger}+b_j^{\phantom{\dagger}})
                  n_{i\sigma}
              +\omega_0\sum_j b_j^{\dagger}b_j^{\phantom{\dagger}}\,.
\label{eqn:hhm}
\end{eqnarray} 
Here $c^{\dagger}_{i\sigma}$ ($c^{}_{i\sigma}$) creates (annihilates)
a spin-$\sigma$ electron at Wannier site $i$ of an 1D lattice 
with $N$ sites, $n_{i\sigma}=c^{\dagger}_{i\sigma}c^{}_{i\sigma}$,
and $b^{\dagger}_{i}$ ($b^{}_{i}$) are the corresponding creation 
(annihilation) operators for a dispersionless phonon.  
We consider the case $\frac{1}{N}\sum_{i\sigma} n_{i\sigma}=1$ hereafter, and take $t$ as energy unit. 

\begin{figure}[t]
  \begin{minipage}{0.38\hsize}
   \begin{center}
    \includegraphics[width=0.95\linewidth]{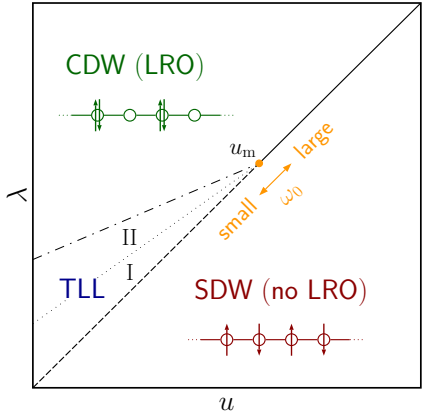}
   \end{center}
  \end{minipage}
  \begin{minipage}{0.615\hsize}
  \begin{center}
     \caption{Qualitative phase diagram of the 1D 
   Hubbard-Holstein model. Given that in the half-filled
   HHM model the ground-state is metallic at $u=0$ for
   $\omega_0>0$ provided that $g<g_c$, it was proposed 
   that this metallic phase continues to exist between
   the SDW and CDW states for $u>0$~\cite{TC03,CH05,TAA05,FHJ08}.
   With increasing $\omega_0$ the region of the intervening metallic state 
   increases, and the tricritical point $u_m$ moves to larger $u$~\cite{CH05}.
   The SDW state shows no long-range order (LRO) and is characterised 
   by a vanishing spin gap $\Delta_s$ but a finite charge gap 
   $\Delta_{c_1}$, whereas the CDW phase exhibits 
   true LRO and $\Delta_s=\Delta_{c_1}>0$.}  
   \label{fig:pdhhm}
  \end{center}
  \end{minipage}
\end{figure}

Based on exact diagonalisation data for the staggered static spin/charge 
structure factor, 
 $S_{\sigma/\rho}(q)=\frac{1}{N}\sum_{j,\ell}e^{iq(j-\ell)}
                     \langle(n_{j\uparrow}\pm n_{j\downarrow})
                            (n_{\ell\uparrow}\pm n_{\ell\downarrow})
                     \rangle$,
it has been argued that the HHM shows a crossover between
Mott and Peierls insulating phases near $u/\lambda\simeq 1$~\cite{FKSW03}.
But this only holds in the strong-coupling adiabatic-to-intermediate
phonon frequency regime. Later on the ground-state phase diagram of the 
HHM was explored in more detail, also for weak interaction strengths
and large phonon frequencies. In this regime, variational displacement 
Lang-Firsov~\cite{TC03}, stochastic series expansion QMC~\cite{CH05}, 
and DMRG~\cite{TAA05}  methods give strong evidence that,
if $\lambda$ is enhanced at fixed $u$ and $\omega_0$, the 
SDW-CDW transition splits into two subsequent SDW-metal and 
metal-CDW transitions at $\lambda_{c1}$ and $\lambda_{c2}$,
respectively (see fig.~\ref{fig:pdhhm}, dashed and dot-dashed lines).
Very recent DMRG data indicated that in the anti-adiabatic regime of
very large phonon frequencies the metallic phase might be even more 
extended than the one obtained by QMC and is subdivided into regions  
with a normal 1D metallic (I) and a bipolaronic-liquid (II) 
behaviour~\cite{FHJ08}.

In this work, we will re-examine the weak-coupling SDW-CDW 
transition regime by calculating the ground-state properties
of the HHM in the framework of a large-scale numerical 
(boson pseudo-site) DMRG approach supplemented by a finite-size 
scaling analysis~\cite{JW98b}.  

Characterising the SDW-CDW-intervening metallic phase of the HHM,  
we presume that for the metallic state a 
Tomonaga-Luttinger-liquid (TLL) description 
holds. In the TLL picture, nonuniversal coefficients, $K_\rho$
and $K_\sigma$, determine the decay of correlation functions
and therefore can be used to identify the properties of the
TLL phase~\cite{Vo95}, but also the phase boundaries to the insulating 
states~\cite{CH05}. 
In practice, we can extract the TLL correlation exponents
from the slope of the corresponding structure factors
in the long-wavelength limit~\cite{CH05,EGN05}: 
$ K_{\rho/\sigma}=\pi\lim_{q\to 0}S_{\rho/\sigma}(q)/q$,
where $q=2\pi/N$ for $N\to \infty$.

Specifically,  $K_{\rho}>1$  ($K_\rho<1$)
corresponds to attractive (repulsive) charge correlations in the TLL  
and $K_\rho=0$ signals an insulating phase. Hence $K_\rho$ jumps 
from $ 1\to 0$ at the metal-SDW/CDW transitions. The spin exponent
takes the value $K_\sigma=0$ in a spin-gapped phase and
$K_\sigma=1$ everywhere else in the thermodynamic limit~\cite{NSO08}.
For finite systems the situation is more involved, in particular
for the spin exponent $K_\sigma$. First,  
the convergence $K_\sigma\to 0$ is slow-going as $N\to \infty$ in 
the spin-gapped phase. Second, logarithmic corrections
prevent $K_\sigma\to 1$ in the spin-gapless (SDW) phase.
On the other hand, these logarithmic corrections vanish at 
the critical point, where the spin gap opens, and we can 
utilise that $K_\sigma$ ($K_{\rho}$) crosses 1 from
above (below) at some $\lambda_{c_1}$ (as the 
electron-phonon coupling increases for fixed $u$),      
in order to determine the SDW-metal phase boundary itself.
Increasing $\lambda$ further, $K_{\rho}$ should cross 1 
once again, this time from above, at another critical 
coupling strength, $\lambda_{c_2}$, which pins the metal-CDW 
transition point down. 

\begin{figure}[t]
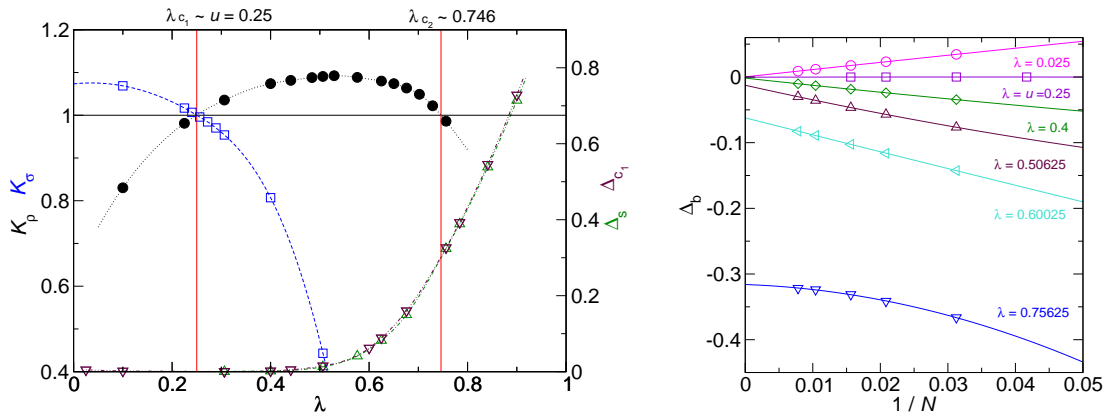

\begin{center}
    \includegraphics[width=0.51\linewidth]{fig1.eps}\hspace*{0.5cm}
   \includegraphics[width=0.35\linewidth]{fig2-2.eps}
 \caption{DMRG results for the half-filled HHM with $u=0.25$ 
and $\omega_0=5$. Left panel: charge ($K_\rho$, filled circles) and 
spin ($K_\sigma$, open squares) TLL exponents as functions of $\lambda$ 
[left-hand axis of ordinate]. Open triangles up (triangles down)
give the spin (charge) gap $\Delta_{s}$ ($\Delta_{c_1}$) 
[right-hand axis of ordinate]. Data shown are extrapolated values
for the infinite systems, using open boundary conditions.
Right panel: Finite-size scaling of the binding energy
$\Delta_b$ for different $\lambda$; lines are polynomial fits.
In the numerical calculations we use up to five pseudo-sites and keep
2400 density-matrix eigenstates; then, for all parameters studied, 
the local boson density is less than $10^{-8}$ and the discarded 
weight is smaller than $10^{-9}$.}
\label{fig2}
\end{center}
\end{figure}

Figure~\ref{fig2} corroborates this scenario for the anti-adiabatic
regime of the HHM. The two critical values $\lambda_{c_1}$ and $\lambda_{c_2}$
are in accord with the phase diagram obtained by QMC~\cite{CH05}.  
$K_\sigma <1$ and $K_{\rho}>1$ earmark the intervening metallic phase.
In terms of the TLL framework, a metallic phase with $K_{\rho}>1$ exhibits
dominant superconducting correlations. Recent DMRG calculations of the
the ($s$--, $p$--, and $d$--wave) superconducting correlation functions 
of the half-filled HHM indicate, however, that these correlations are only 
sub-dominant against CDW correlations~\cite{TAA05}, while QMC 
investigations attributed the $K_{\rho}>1$ to finite-size effects 
and suggest that $K_{\rho}(N\to \infty) =1$, i.e., superconducting and 
CDW correlations are exactly degenerate.  

Here we inspect the finite-size scaling of the spin and single-particle
charge excitation gaps, 
$\Delta_{s}(N)=E_0(1)-E_0(0)$ and   
$\Delta_{c_1}(N)=E_0^{+}(1/2)+E_0^{-}(-1/2)-2E_0(0)$,
respectively, as well as that of the two-particle binding energy
$\Delta_{b}(N)=E_0^{2-}(0)+E_0(0)-2 E_0^{-}(-1/2)$,
where $E_0^{(L\pm)}(S^z)$ is the ground-state 
energy at or away from half-filling with $N_{e}=N\pm L$  
particles in the sector with total spin-$z$ component $S^z$. 
The left panel of fig.~\ref{fig2} shows
that both spin and charge gaps open at $\lambda_{c_1}$ (but  
there is no LRO). For $u<u_m$, the transition at $\lambda_{c_1}$ seems to be of 
Kosterlitz-Thouless type, i.e. just above $\lambda_{c_1}$ the gaps 
are exponentially small and therefore their magnitude is difficult 
to determine. In this region, denoted by (I) 
in fig.~\ref{fig:pdhhm}, we find $\Delta_{c_1}\sim\Delta_{s}$, and 
the binding energy $\Delta_{b}$ is also extremely small, or maybe even 
zero (see triangles up, right-hand panel of fig.~\ref{fig2}). As $\lambda$
increases, we obtain a (smooth) crossover to a metallic regime with 
a noticeable two-particle binding $\Delta_{b}<0$ 
(region (II) in fig.~\ref{fig:pdhhm}),
where $\Delta_{c_1}\sim\Delta_{s}$. This is in accord with the very recent  
findings of Ref.~\cite{FHJ08}, where a subdivision of the metallic phase
into a weakly renormalised TLL (I) and a bipolaronic liquid\footnote{Note 
that the polaronic two-particle bound states are not necessarily small
(i.e. on-site).} (II) 
was suggested. In the latter phase, the two-particle excitation gap
$\Delta_{c_2}(N)=E_0^{2+}(0)+E_0^{2-}(0)-2 E_0(0)$ was shown to scale
to zero. In the CDW phase, which typifies a bipolaronic superlattice
at large phonon frequencies, we have, besides 
$\Delta_{s}=\Delta_{c_1}> 0$, $\Delta_{c_2}>0$ and 
$\Delta_{b}<0$, whereas in the SDW state $\Delta_{c_2}>0$ but 
$\Delta_{b}(N\to\infty)\to 0$. While the basic scenario discussed
so far persists in the adiabatic regime, the metallic region shrinks 
as the phonon frequency $\omega_0$ becomes smaller~\cite{CH05,FHJ08}.  
Furthermore, the CDW state rather behaves as a normal Peierls insulator and
consequently there is a weaker tendency towards bipolaron formation
in the metallic state for small $\lambda$, and $u<u_m$.    

Finally, let us investigate the behaviour of the momentum distribution 
function, 
$n_\sigma(k) =\frac{1}{N}\sum_{j,l=1}^{N}\cos \left(k(j-l)\right)
         \langle 
 	 c^\dagger_{j,\sigma} c^{\phantom{\dagger}}_{l,\sigma}
         \rangle$,
$k=2\pi m /N$, $m=0,\dots,N/2$,
which can be obtained by DMRG for a system with periodic 
boundary conditions. 

\begin{figure}[t]
  \begin{minipage}{0.52\hsize}
   \begin{center}
    \includegraphics[width=0.9\linewidth]{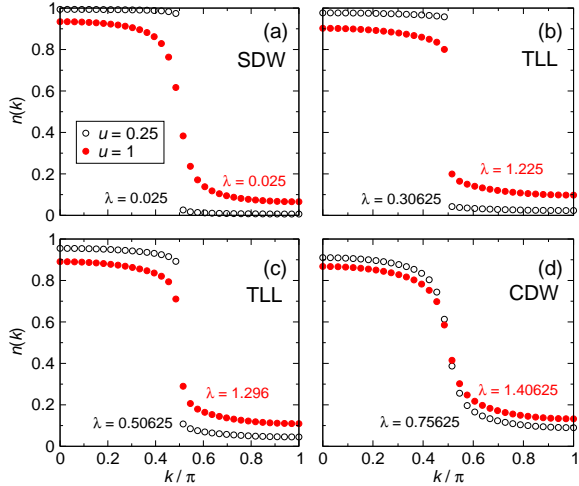}
   \end{center}
  \end{minipage}
  \begin{minipage}{0.475\hsize}
  \begin{center}
   \caption{Momentum distribution of the half-filled HHM
   in the anti-adiabatic regime ($\omega_0=5$).  Open (closed) circles 
   give DMRG results at $u=0.25$ ($u=1$) for a system with $N=66$ sites 
   and periodic boundary conditions. The occupation of fermionic states 
   carrying momentum $k$ is given by
   $n(k)=\frac{1}{2}\sum_\sigma n_\sigma(k)$, and we have $k_{F}=\pi/2$,  
   $n_{k_{F}}=1/2$ for the half-filled band case. 
   In the intermediate metallic phase, $n(k)$ exhibits a 
   power-law singularity at $k_F$ [see panels (b) and (c)].
   At weak and strong electron-phonon couplings insulating SDW 
   [panel (a)] and CDW [panel (d)]
   are realised, respectively.}
   \label{fig:nk}
  \end{center}
  \end{minipage}
\end{figure} 

Figure~\ref{fig:nk} shows the variation of $n(k)$ for weak (circles) and 
intermediate (stars) Hubbard interactions in the SDW (a), TLL (b)-(c), 
and CDW (d) phases.  The momentum distribution is a monotonously decreasing 
function as $k$ changes from the centre ($k=0$) to the boundary 
of the Brillouin zone ($k=\pi$). Since we consider the weak-coupling regime,
$n(k)$ is only weakly renormalised away from the Fermi momentum $k_F$. 
For a 1D TLL, instead of the Fermi liquid typical jump of $n(k)$ at $k_F$, 
one finds an essential power-law singularity~\cite{Vo95}, corresponding 
to a vanishing quasiparticle weight $Z=0$. For finite TLL systems, 
the difference $\Delta=n(k_{F}-\delta)-n(k_{F}+\delta)$ 
is finite (with $\delta=\pi/N=\pi/66$ in our case), and
rapidly decreases with increasing couplings $\lambda$, $u$. 
Approaching the insulating SDW/CDW states $n(k)$ becomes 
a smooth curve, i.e. the singularity vanishes and $\Delta\to 0$. 
At very large $\lambda$, the system develops a ``perfect'' CDW 
with $n(k)=1/2$ for all momenta $k$.  

To summarise, we validated the existence of an intervening metallic
phase in the SDW-CDW transition regime of the 1D half-filled 
Hubbard-Holstein model for $u<u_m$ by large-scale DMRG calculations. 
Spin and charge gaps open exponentially slowly at the SDW-TLL 
transition point, $\lambda_{c_1}$, but no long-range order develops.
$\lambda_{c_1}(u,\omega_0)$ can be determined from the ``1''--crossing of the 
spin and charge TLL parameters. In the TLL, the momentum distribution 
function exhibits a power-law singularity at $k_F$. 
As the electron-phonon coupling increases, a crossover
to a bipolaronic metal, indicated by negative binding energy,
takes place, before the systems enters the long-range ordered
insulating CDW phase at a second critical coupling $\lambda_{c_2}$. 
We would like to point out that fixing the metal-CDW phase boundary 
quantitatively is a difficult issue.

{\it Acknowledgements.} The authors would like to thank  G. Hager
and E. Jeckelmann  
for valuable discussions. 
This work was supported by KONWIHR Bavaria, and DFG through SFB 652.


\end{document}